\documentclass{jltp}

\usepackage{graphicx}
\begin{document}
\title{Study of the Order Parameter symmetry in overdoped Y$_{1- 
x}$Ca$_{x}$Ba$_{2}$Cu$_{3}$O$_{7-\delta}$ by measurement of 
Andreev reflection, evidence for enhancement of the sub dominant order 
parameter amplitude in high transparency contacts}

\author{Amir Kohen and Guy Deutscher}

\address{School of Physics and Astronomy, Raymond and Beverly Sackler\\
  Faculty of Exact Science, Tel Aviv University Ramat Aviv 69978, Israel}

\runninghead{A. Kohen and G. Deutscher}{Study of the Order Parametr 
symmetry in overdoped Y$_{1-x}$Ca$_{x}$Ba$_{2}$Cu$_{3}$O$_{7-\delta}$}
\maketitle
\begin{abstract}
We have measured the differential conductance of Au/Y$_{1 -
  x}$Ca$_{x}$Ba$_{2}$Cu$_{3}O_{7-\delta}$ point contacts in the regime 
where it is dominated by Andreev reflection, which enhances its value at
low bias. We find that the characteristics can not be fitted by a pure
d-wave Order Parameter (OP). Using the formalism developed by Kashiwaya
and Tanaka\cite{Kashiwaya:1995}, the best fits are obtained by adding a
sub-dominant
imaginary OP, whose amplitude appears to depend on the transparency of the
contact. At high transparencies it can reach up to 60{\%} of the total amplitude of the OP, for lower transparencies it is substantially smaller. We attribute this enhancement, at high transparencies to a proximity effect between the Au tip and the superconducting electrode.

\end{abstract}

\section{INTRODUCTION}

Determining the symmetry of the order-parameter (OP) in high temperature 
superconductors (HTS) has been the subject of both experimental and 
theoretical studies. In the case of YBaCuO it was found that the OP is 
dominated by d-wave symmetry \cite{Tsuei:2000}. One of the important 
experimental results in support of this was the zero bias conductance peak 
(ZBCP) measured in tunneling experiments \cite{Covington:1997}. Dagan et 
al.\cite{Dagan:2001} have shown, experimentally, that in over-doped, 
(110) oriented, YBCO samples a spontaneous splitting of the ZBCP occurs at 
low temperatures. They have related this splitting to the appearance of a 
sub-dominant imaginary OP (SIOP), with a maximum value of around 25{\%} of 
the d-wave amplitude. Sharoni et al.\cite{Sharoni:2002} using STM 
junctions have found a similar behavior in YCaBaCuO, (110) oriented, 
over-doped samples. Both results were obtained in N/I/S contacts. Tanuma 
et al.\cite{Tanuma:2001} have shown that in the case of a (110) surface of a 
$d_{x^{2}-y^{2}}$ superconductor (SC) in contact with a normal metal, 
the amplitude of the d-wave OP is reduced in the vicinity of the interface. 
This reduction allows for the appearance of a SIOP of either \textit{id}$_{xy}$ or 
\textit{is}$_{ }$ symmetry. They have further shown that the larger is the
reduction in 
the $d_{x^{2}-y^{2}}$ amplitude the larger is the SIOP amplitude. Y. 
Ohashi\cite{Ohashi:1996} have predicted a proximity effect, where a 
small, s-wave symmetry OP, is induced in a normal metal by a SC with a 
$d_{x^{2}-y^{2}}$ symmetry OP. This effect appears in the case of a 
normal metal and a (100) surface of a $d_{x^{2}-y^{2}}$ SC and is 
enhanced in high transmission contacts. The appearance of the induced s-wave 
OP in the normal metal causes a reduction of the d-wave amplitude near the 
N/S interface in the SC side and the higher is the amplitude of the s-wave 
OP, in the normal metal, the larger is the reduction in the
$d_{x^{2}-y^{2}}$ amplitude. In this work we report on N/S, point contact
differential conductance measurements. We have analyzed our results using
the formalism of Kashiwaya et al. \cite{Kashiwaya:1995}. We find, in
agreement with the results of 
Ref\cite{Dagan:2001,Sharoni:2002,Farber:2002}, that the curves are best 
fitted by a dominant d-wave OP with an additional SIOP. Our results are best 
fitted with a SIOP having a \textit{s-wave }symmetry leading to an OP of 
$d_{x^{2}-y^{2}}$\textit{+is} symmetry. The maximum amplitude of the SIOP reaches 60{\%} of the OP 
amplitude, which is considerably higher in comparison to previously reported 
values. We attribute this enhancement to a proximity effect between the HTS 
film and the Au electrode. Our results are best fitted with low Z values 
($0.3<Z<0.7$), where $Z = \frac{H}{\hbar v_F }$ is a dimensionless parameter, 
defined by Bolnder et al. \cite{Blonder:1982}, proportional to the 
strength of the barrier at the interface. This indicates that we have 
produced low reflectivity N/S contacts and thus we expect the proximity 
effect to be significant. We find that the amplitude of the SIOP is indeed a 
decreasing function of Z.

\section{EXPERIMENTAL}

Y$_{1 - x}$Ca$_{x}$Ba$_{2}$Cu$_{3}$O$_{7 - \delta }$ samples with 
x=0.05,0.1 and 0.2 were grown by off-axis sputtering. We have used the 
growth conditions described in Ref \cite{Sharoni:2002} to produce, as 
determined by x-ray diffraction, (100), (110) and (001) orientated samples. 
Sharoni et al.\cite{Sharoni:2002}, using STM topographic scans have 
shown that (110) films grown under the same conditions expose (100) facets. 
SEM and AFM pictures of the (001) films show a-axis grains on the surface of 
the film. Thus, in all three cases (100), (110) and (001) films, (100) 
facets are exposed at the film surface. Ozawa et al.\cite{Ozawa:1997} 
have studied the electronic properties of the surfaces of (110) oriented 
YBCO films. They have found that the degradation time of (110) facets is 
significantly smaller than that of (100) facets. We therefore expect that 
the chances of obtaining a metallic contact with a (100) facet are 
considerably higher in comparison to a (110) facet in our films. 
The contacts were formed using a mechanically cut Au tip mounted on a 
differential screw. The measurements were taken in a liquid Helium Dewar 
at a temperature of 4.2 K. I/V characteristics were measured digitally and
differentiated numerically to yield the differential conducatnce as a
function of voltage. Each I/V measurement comprises of two sweeps to check for the absence of 
heating histerisis effects 

\section{RESULTS AND DISCUSSION}

Figure 1. shows the result of one of the contacts. One can easily detect two 
distinct energy scales, see arrows in the figure. Indeed the experimental 
curve was best fitted with the $d_{x^{2}-y^{2}}$\textit{+is }symmetry where 
$\Delta _d $=15.9 meV, $\Delta _S $=8.5 meV and Z=0.46. The other fit 
parameters are $\alpha = 0$ (corresponding to a (100) contact),
$\Gamma=$1.7meV
(representing inelastic scattering) and T=4.2K. We have performed 
the integration over the direction of injected electrons from
$-\frac{\pi}{2}$
to $\frac{\pi }{2}$ with equal weight. Using the same symmetry of the 
OP we were able to fit all other 9 contacts by adjusting the values
of values of $\Delta_{d},\Delta_{s}$, Z and $\Gamma$. Some of the curves
can be fitted also using a 
$d_{x^{2}-y^{2}}$\textit{+id}$_{xy}$ symmetry OP, however this requires
using a narrow 
tunneling cone in the fit, which is unreasonable for a metallic, low 
barrier, contact. As for the $d+s$ OP symmetry, it requires using
$\Delta_{s}>\Delta_{d}$, thus, we find it less probable, as it 
fails to explain experimental data obtained for high Z contacts 
\cite{Dagan:2001,Sharoni:2002} and would suggest that the 
s-wave channel is stronger than the d-wave one, which is in conflict with 
the findings of most experimental data \cite{Tsuei:2000}. Figure 2 
summarizes our results for the different contacts. As the films have 
different critical temperatures, we have plotted the normalized amplitudes
of 
the OP components,
$\frac{\Delta_{s}}{\sqrt{\left(\Delta_{s}\right)^{2}+\left(\Delta_{d}\right)^{2}}}\equiv\delta_{s}$
and $\frac{\Delta_{d}}{\sqrt{\left(\Delta_{s}\right)^{2}+\left(\Delta_{d}\right)^{2}}}\equiv\delta_{d}$
as a function 
of Z. Fig. 2. shows, that the relative amplitude of the \textit{is} component increases 
from values as low as 0-10{\%} for z around 0.7 up to values of around 60 
{\%} for Z around 0.3. At the same time the relative amplitude of the
d-wave 
component decreases. The slope of $\delta_{s}$  as a function of Z is
maximum at Z$\simeq$0.5. We see a 
change from an OP which is an almost pure d-wave for the high Z contacts to 
an OP, which has almost equal amplitudes of the two components in the low Z 
regime. To conclude we have shown that our data corresponds to a 
$d_{x^{2}-y^{2}}$\textit{+is }symmetry, with the value of the SIOP being a 
decreasing function of Z. The enhancement of the \textit{is }component, for high 
transparency contacts, leads us to conclude that  it is a result of a special 
proximity effect between the normal metal electrode and the d-wave SC, which 
is much larger than the one predicted by Y. Ohashi\cite{Ohashi:1996}. 
We can explain our data in the following way. As predicted by 
Y.Ohashi\cite{Ohashi:1996}, a normal metal in contact with a (100) 
boundary of a d-wave SC results in a decrease of the d-wave near the 
boundary and the appearance of an s-wave symmetry OP in the normal metal. 
This results in a loss of condensation energy on the SC side. However if an 
\textit{is }OP develops in the d-wave SC in the vicinity of the barrier, as predicted by 
Tanuma et al.\cite{Tanuma:2001} for a (110) oriented contact, we get a 
state, which is energetically favorable and still matches the s-wave OP on 
the normal metal side. Therefore we conclude that the d-wave SC has a 
subdominant s-wave pairing channel, as suggested by Fogelstr\"{o}m et 
al.\cite{Fogelstrom:1997} , otherwise the appearance of the \textit{is} OP would not have been energetically favorable.

\begin{figure}[ht]
\centerline{\includegraphics[height=3.2in]{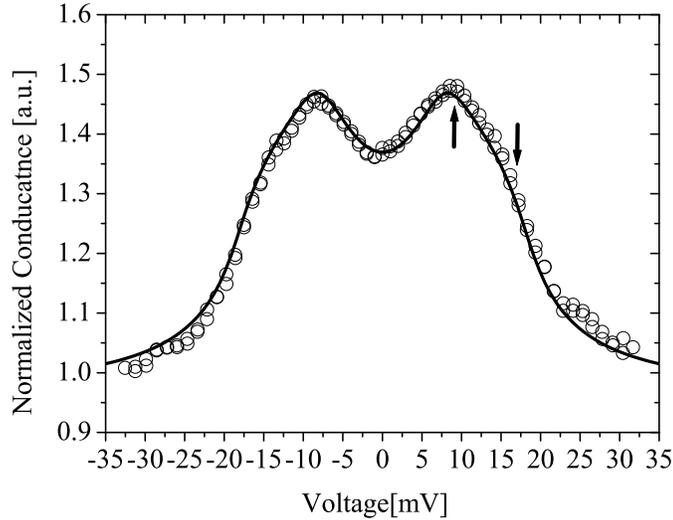}}
\caption{Y$_{0.8}$Ca$_{0.2}$Ba$_{2}$Cu$_{3}$O$_{7-\delta}$/Au
  contact. Normalized conduductance vs. voltage,
  T$=4.2$K, R$_{N}$=12.5$\Omega$. (circles). Fit using
  $\Delta_{d}=$15.9meV, $\Delta_{s}$=8.5meV, Z=0.46,
  $\Gamma=$1.7meV, (100) contact, (line). Arrows point out to the
  manifestation of the two energy scales, peak and change of slope.}
\label{fig1}
\end{figure}
\begin{figure}[ht]
\centerline{\includegraphics[height=3.2in]{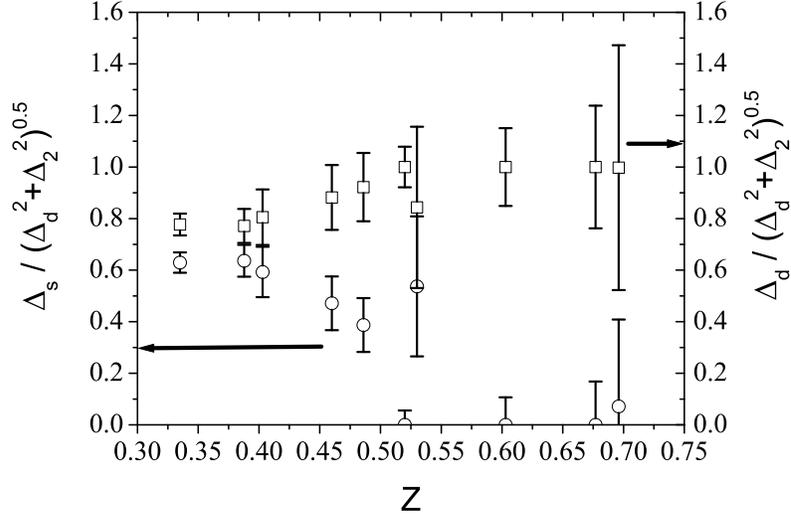}}
\caption{The relative weight of the OP components, $\delta_{s}$
(circles) and $\delta_{d}$ (squares) as a function of the barrier
  strength parameter, Z.}
\label{fig2}
\end{figure}

\section*{ACKNOWLEDGMENTS}

We are indebted to Roy Beck for the computerized implementation of the 
algorithm used to fit the data. This work was supported in part by the 
Heinrich Hertz-Minerva Center for High Temperature Superconductivity, and by 
Oren Family chair of Experimental Solid State Physics.


\begin{thebibliography}{9}
\bibitem{Kashiwaya:1995}S. Kashiwaya et al. Phys. Rev. B \textbf{51} 1350, (1995)
\bibitem{Tsuei:2000}C.C. Tsuei and J.R. Kirtley, Rev. Mod. Phys. \textbf{72}, 969 (2000)
\bibitem{Covington:1997}M. Covington et al. Phys. Rev. Lett. \textbf{79, }277 (1997)
\bibitem{Dagan:2001}Y. Dagan and G.Deutscher Phys. Rev . Lett. \textbf{87} 177004, (2001)
\bibitem{Sharoni:2002}A. Sharoni et al. Phys. Rev. B \textbf{65}, 134526 (2002)

\bibitem{Tanuma:2001}Y. Tanuma and Y.Tanaka Phys. Rev. B \textbf{64},
214519(2001)
\bibitem{Ohashi:1996}Y. Ohashi J. Phys. Soc. Jpn. \textbf{65}, 823 (1996)
\bibitem{Blonder:1982}G.E. Blonder et al. Phys. Rev. B \textbf{25}, 4515 (1982).
\bibitem{Ozawa:1997}Ozawa et al. IEEE Trans. On Applied Superconductivity. \textbf{7} (1997)
\bibitem{Farber:2002}E. Farber and G. Deutscher  these proceedings
\bibitem{Fogelstrom:1997}M. Fogelstr\"{o}m et 
al. Phys. Rev. Lett. \textbf{79}, 281(1997) 
\end{thebibliography}
\end{document}